\documentclass[12pt]{iopart}
\usepackage{graphicx}
\usepackage{epsfig}
\usepackage{soul}
\usepackage{iopams}
\begin{document}

\title{Distillability sudden death in qutrit-qutrit systems under amplitude damping}
\author{Mazhar Ali}
\address{Fachbereich Physik, Universit\"{a}t Siegen, 57068, Germany}
\ead{mazharaliawan@yahoo.com}

\begin{abstract}
Recently it has been discovered that certain two-qutrit entangled states interacting with global and/or multi-local decoherence undergo distillability sudden death (DSD). We investigate this phenomenon for qutrit-qutrit systems interacting with statistically independent zero-temperature reservoirs. We show that certain initially prepared free-entangled states become bound-entangled in a finite time due to the action of Markovian dissipative environment. Moreover, in contrast with local dephasing, simple local unitary transformations can completely avoid distillability sudden death under amplitude damping.
\end{abstract}

\submitto{\JPB}
\maketitle

\section{Introduction}

Quantum entanglement is a key resource for quantum information sciences \cite{Nielsen2000}. As the quantum states interact with their environment therefore it is very important to understand how an initial entanglement is affected by this interaction. The dynamical aspects of entanglement are essential not only for the utilization of entangled states in quantum technology but for a deep understanding of the transition from quantum to classical world as well. Recently, Yu-Eberly \cite{YE-PRB-2003, Yu-Eberly} investigated the dynamics of two-qubit entanglement under various types of decoherence. They found that although the coherence of a single qubit may take infinite time to vanish, nevertheless entanglement can disappear at a finite time, a phenomenon named as ``entanglement sudden death" (ESD). Clearly, such finite-time disentanglement can seriously affect applications of entangled states in quantum information processing. The initial report of ESD in two-qubit entangled states in a specific model was later explored in wider contexts and in higher dimensions of Hilbert space where the qubits are replaced by qutrits (three dimensional systems) or qudits ($d$ dimensional systems) \cite{ESD-2007, ESD-2008, Ann-Jaeger, Rau, ESD-2009}. The experimental evidences for this effect have been reported for optical setups \cite{Exp1-ESD} and atomic ensembles \cite{Exp2-ESD}. Recently, it has been observed that \cite{ESB-2008} if there is entanglement sudden death in the principal system then there is always entanglement sudden birth (ESB) in the reservoirs. Interestingly, such entanglement sudden birth may appear earlier, simultaneously or later than entanglement sudden death.

Entanglement of qubit-qubit and qubit-qutrit systems has been completely characterized by Peres-Horodecki criterion \cite{Pe-PRL96, HHH-PLA96}. This criterion would imply that if a quantum state in Hilbert space of qubit-qubit and qubit-qutrit systems has positive partial transpose (PPT), then it is separable. For higher dimensional bipartite systems such characterization is not an easy task. Bipartite entangled states are divided into free-entangled states and bound-entangled states \cite{Horodecki-RMP-2009, Horodecki-PRL80-1998}. Free-entangled states can be distilled under local operations and classical communication (LOCC) whereas bound-entangled states can not be distilled to pure-state entanglement no matter how many copies are available. It was observed that bound entanglement may activate teleportation fidelity \cite{Horodecki-PRL82-1999} and manifests the irreversibility in asymptotic manipulation of entanglement \cite{Yang-PRL95-2005}. Bound entanglement constructed by purely mathematical arguments have existence in physical processes \cite{Toth-PRL99-2007}.

Recently, it was proposed that free-entangled states may be converted into bound-entangled states under either multi-local decoherence \cite{Song-PRA80-2009} and/or under combination of collective and local dephasing processes \cite{Ali-arXiv2009}. In particular, it was shown that certain free-entangled states of qutrit-qutrit systems become non-distillable in a finite time under the influence of classical noise. Such behavior has been named as distillability sudden death (DSD) \cite{Song-PRA80-2009}. This discovery raises an interesting question that whether such behavior is restricted to a certain type of environment or it is also a generic feature like entanglement sudden death. We have studied this phenomenon for certain qutrit-qutrit states which undergo the simple process of spontaneous emission. We have analyzed the dissipative dynamics of three-level V-type atomic system where spontaneous emission may take place from two excited levels to the ground state and direct transition between excited levels is not allowed. We have found that distillability sudden death may occur in this type of open systems as well. This suggests that DSD might be a generic feature such that free-entangled quantum states dynamically convert to bound-entangled states. However, we have noticed that unlike local and global dephasing processes, simple local unitary transformations can avoid this phenomenon in our specific model.

This paper is organized as follows. In Sec. \ref{Sec:Model}, we discuss the physical model and the basic equation of motion along with its solution. In Sec. \ref{Sec:Results}, we discuss briefly the idea of distillability sudden death and demonstrate the possibility of distillability sudden death under amplitude damping. We show that locally equivalent quantum states do not exhibit such behavior. We discuss the physical reason for distillability sudden death in a particular family of quantum states. We conclude our work in Sec. \ref{Sec:Conc}.

\section{Dynamics of qutrit-qutrit system under amplitude damping} \label{Sec:Model}

\subsection{Three-level atom and quantum interference}

\begin{figure}[h]
\scalebox{2.0}{\includegraphics[width=1.8in]{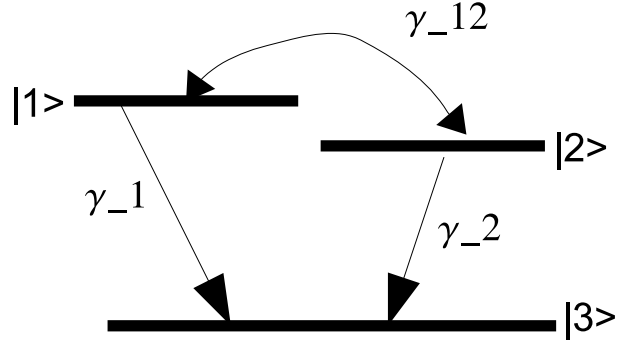}}
\centering
\caption{Three level atom is shown in $V$ configuration as system-$\mathrm{I}$.}
\label{Fig:V-conI}
\end{figure}

We consider a three-level atom in the $V$ configuration. Let $|1\rangle$, $|2\rangle$ be the two nondegenerate excited states with transition frequencies to the ground state $|3\rangle$ represented by $\omega_1$, $\omega_2$ and electric dipole moments $\vec{\mu}_1$, $\vec{\mu}_2$ respectively. We assume that excited states can decay to the ground state by spontaneous emission and a direct transition between excited states is not allowed. The spontaneous decay constants of the excited states $|j\rangle$, $j = 1,\,2$ are given as 
\begin{eqnarray}
\gamma_j = \frac{2 \, |\vec{\mu}_j|^2}{3 \, \hbar } \, \bigg(\frac{\omega_j}{c} \bigg)^3 \, .
\end{eqnarray}
It is well known that if the dipole moments of these two transitions are parallel, then an indirect coupling between levels $|1\rangle$ and $|2\rangle$ can appear due to interaction with vacuum \cite{Agarwal-Book}. This coupling between transitions $|1\rangle \, \to \, |3\rangle$ and $|2\rangle \, \to \, |3\rangle$ is termed as quantum interference with cross damping constant
\begin{eqnarray}
\gamma_{12} = \beta_{\mathrm{I}} \, \sqrt{\gamma_1 \, \gamma_2} \, ,
\end{eqnarray}
where
\begin{eqnarray}
\beta_\mathrm{I} = \frac{\vec{\mu}_1 \cdot \vec{\mu}_2}{|\vec{\mu}_1| \, |\vec{\mu}_2|} \, .
\end{eqnarray}
The parameter $\beta_\mathrm{I}$ represents the mutual orientation of transition dipole moments such that for maximum quantum interference $\beta_\mathrm{I} = 1$ and for $\beta_\mathrm{I} = 0$ it (interference) vanishes. Figure \ref{Fig:V-conI} shows this typical configuration.

The time evolution of this system-$\mathrm{I}$ is given by the master equation \cite{Ficek-PRA69-2004, Derkacz-PRA74-2006} 
\begin{eqnarray}
\frac{d \, \rho}{dt} = - \, i \, [H \, , \, \rho] + \Lambda_\mathrm{I} \, \rho \, , \label{Eq:MESI}
\end{eqnarray}
where the damping term is given as
\begin{eqnarray}
\Lambda_\mathrm{I} \, \rho &=& \frac{\gamma_1}{2} \, ( 2 \, \sigma_{31} \, \rho \, \sigma_{13} - \sigma_{11} \, \rho - \rho \, \sigma_{11}) + \frac{\gamma_2}{2} \, ( 2 \, \sigma_{32} \, \rho \, \sigma_{23} \nonumber \\ && - \sigma_{22} \, \rho - \rho \, \sigma_{22}) + \frac{\gamma_{12}}{2} \, ( 2 \, \sigma_{31} \, \rho \, \sigma_{23} - \sigma_{21} \, \rho - \rho \, \sigma_{21}) \nonumber \\&&  + \frac{\gamma_{12}}{2} \, ( 2 \, \sigma_{32} \, \rho \, \sigma_{13} - \sigma_{12} \, \rho - \rho \, \sigma_{12})\, . \label{Eq:DSI}
\end{eqnarray}
Here $\sigma_{kl}$ is the atomic transition operator taking an atom from level $|l\rangle$ to $|k\rangle$. As we are interested in the time evolution of initial states due to the spontaneous emission, we can ignore the Hamiltonian part (for details see \cite{Ficek-PRA69-2004, Derkacz-PRA74-2006}).

\begin{figure}[h]
\scalebox{2.0}{\includegraphics[width=1.8in]{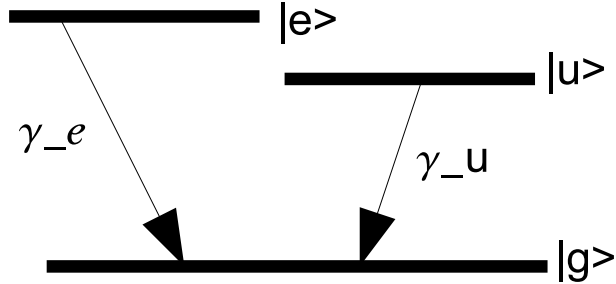}}
\centering
\caption{Three level atom (system-$\mathrm{II}$) is shown in $V$ configuration. Two excited states $|e \rangle$ and $|u\rangle$ are allowed to make transitions to ground state $|g\rangle$ with decay rate $\gamma_e$ and $\gamma_u$, respectively. The transition between excited states is dipole forbidden.}
\label{Fig:V-con}
\end{figure}
In atomic spectroscopy, the transition dipole moments are usually perpendicular and system-$\mathrm{I}$ described by Eq.(\ref{Eq:DSI}) is difficult to realize. However, it was shown \cite{Ficek-PRA69-2004} that the effects of quantum interference can be simulated to a large degree by three-level system-$\mathrm{II}$ as shown in Figure \ref{Fig:V-con} with excited states $|e\rangle$, $|u\rangle$ and the ground state $|g\rangle$. The corresponding master equation contains the damping operator, 
\begin{eqnarray}
\Lambda_\mathrm{II} \, \rho  &=& \frac{\gamma_e}{2} \, ( 2 \, \sigma_{ge} \, \rho \, \sigma_{eg} - \sigma_{ee} \, \rho - \rho \, \sigma_{ee}) + \frac{\gamma_u}{2} \, ( 2 \, \sigma_{gu} \, \rho \, \sigma_{ug} \nonumber \\ && - \sigma_{uu} \, \rho - \rho \, \sigma_{uu}) \, . \label{Eq:DSII}
\end{eqnarray}
The main arguments from Ref.~\cite{Ficek-PRA69-2004} supporting this idea are sketched as follows. We consider the system-$\mathrm{I}$ with $\gamma_1 = \gamma_2 = \gamma$. By taking symmetric and antisymmetric superpositions of the excited states
\begin{eqnarray*}
|s\rangle = \frac{1}{\sqrt{2}} \, ( | 1 \rangle + | 2 \rangle )\, , \quad |a\rangle = \frac{1}{\sqrt{2}} \, ( | 1 \rangle - | 2 \rangle )\, ,
\end{eqnarray*}
and the ground state $|3\rangle$ as a new basis in $\mathbb{C}^3$, the damping term (\ref{Eq:DSI}) becomes
\begin{eqnarray}
\tilde{\Lambda}_\mathrm{I} \, \rho &=& \frac{\gamma}{2} (1 + \beta_\mathrm{I}) (\sigma_{3s} \, \rho \, \sigma_{s3} - \sigma_{ss} \, \rho - \rho \, \sigma_{ss}) + \frac{\gamma}{2} (1-\beta_\mathrm{I}) \nonumber \\ && \times (\sigma_{3a} \, \rho \, \sigma_{a3} - \sigma_{aa} \, \rho - \rho \, \sigma_{aa}) \, . \label{Eq:NB}
\end{eqnarray}
We note that for almost parallel dipole moments $\beta_\mathrm{I} \simeq 1$, the antisymmetric level is metastable. The system-$\mathrm{II}$ with damping operator (\ref{Eq:DSII}) is a generalization of this system in which we allow the two excited states $|e\rangle$ and $|u\rangle$ to be nondegenerate with arbitrary decay rates to the ground state $|g\rangle$. Moreover, the dipole moments can be perpendicular. A measure of quantum interference is defined \cite{Ficek-PRA69-2004, Derkacz-PRA74-2006} as 
\begin{eqnarray}
\beta_\mathrm{II} = \frac{\gamma_e - \gamma_u}{\gamma_e + \gamma_u} \,.
\end{eqnarray}
This expression shows that maximum interference are expected in system-$\mathrm{II}$ when the level $|u\rangle$ is metastable. It was demonstrated \cite{Ficek-PRA69-2004} that system-$\mathrm{II}$ simulates quantum interference in a three-level atom. Moreover, it is simpler model and can be studied analytically. We focus on this model and concentrate on the question that whether certain free-entangled states undergo distillability sudden death or not. As the effects of quantum interference on disentanglement have been studied \cite{Derkacz-PRA74-2006}, we would just take a fixed value of interference parameter and study our question of interest in what follows.

\subsection{Time evolution of pair of three-level system}

We consider the system of two three-level atoms $A$ and $B$ both of type ($\mathrm{II}$) such that they are separated by a distance which is much larger than wavelength of radiation. This condition implies that both atoms are independently interacting with their own local reservoirs. This case is simpler to analyze and can be solved exactly. The two atoms may be initially prepared in an entangled state. The master equation describing the dissipative part of dynamics is given as \cite{Derkacz-PRA74-2006}
\begin{eqnarray}
\frac{d\, \rho}{dt} = \Lambda \, \rho \, , \label{Eq:ME}
\end{eqnarray}
with
\begin{eqnarray}
\Lambda \, \rho =& \frac{\gamma_e}{2} (2 \, \sigma_{ge}^A \, \rho \, \sigma_{eg}^A - \sigma_{ee}^A \, \rho - \rho \, \sigma_{ee}^A) + \frac{\gamma_u}{2} (2 \, \sigma_{gu}^A \, \rho \, \sigma_{ug}^A \nonumber \\& - \sigma_{uu}^A \, \rho - \rho \, \sigma_{uu}^A) + \frac{\gamma_e}{2} (2 \, \sigma_{ge}^B \, \rho \, \sigma_{eg}^B - \sigma_{ee}^B \, \rho - \rho \, \sigma_{ee}^B) \nonumber \\& + \frac{\gamma_u}{2} (2 \, \sigma_{gu}^B \, \rho \, \sigma_{ug}^B - \sigma_{uu}^B \, \rho - \rho \, \sigma_{uu}^B)\,,
\end{eqnarray}
and where
\begin{eqnarray}
\sigma_{ij}^A = \sigma_{ij} \otimes \mathbb{I}_3 \,, \quad \sigma_{ij}^B = \mathbb{I}_3 \otimes \sigma_{ij} \, , \quad i,j = e,u,g.
\end{eqnarray}
The spontaneous emission of atoms $A$ and $B$ from their excited states $|e\rangle$ to their ground states $|g\rangle$ is described by the spontaneous decay rate $\gamma_e$ whereas $\gamma_u$ is the spontaneous decay rate of excited states $|u\rangle$ to ground states $|g\rangle$. The atomic transition operator $\sigma_{ij} = |i \rangle\langle j|$ takes an atom from its state $|j \rangle$ to state $|i \rangle$. Let us identify $|2\rangle$, $|1\rangle$, and $|0\rangle$ be the first excited state, second excited, and ground state of the qutrit, respectively. We choose the basis $ \{ \, |2,2\rangle$, $|2,1\rangle$, $|2,0\rangle$, $|1,2\rangle$, $|1,1\rangle$, $|1,0\rangle$, $|0,2\rangle$, $|0,1\rangle$, $|0,0\rangle \, \}$. The most general solution of the master equation (\ref{Eq:ME}) has been provided for matrix elements of any initial density matrix in Appendix $A$. From this general solution, it is obvious that for $\gamma_u > 0$, all initial states dynamically reduce to ground state of both atoms at sufficiently large time ($t \to \infty$), which is a separable state i.\,e. $|0,0\rangle$. For maximum interference ($\gamma_u = 0$), there may be some asymptotic entangled states \cite{Derkacz-PRA74-2006}. However, we are only interested in the dynamics with $\gamma_u > 0$.

\section{Distillability sudden death} \label{Sec:Results}

First we discuss the characterization of bound-entangled states and then demonstrate the existence of distillability sudden death. It was shown \cite{Horodecki-PRL80-1998} that bound-entangled states have positive partial transpose (PPT) and therefore non-distillable under LOCC. In general, there is not a unique criterion to detect all bound-entangled states. Even for qutrit-qutrit system, there are various bound-entangled states and a single criterion is not capable to detect all of them (see for example \cite{Clarisse} and references therein). However, one can use the realignment criterion \cite{Chen-QIQ-2003} to detect certain bound-entangled states. The realignment of a given density matrix is obtained as $(\rho^R)_{ij,kl} = \rho_{ik,jl}$. For a separable state $\rho$, realignment criterion implies that $\|\rho^R\| \leq 1$. For a PPT-state, the positive value of the quantity $\|\rho^R\|-1$ can prove the bound-entangled state.

Distillability sudden death corresponds to a situation where free-entangled states become non-distillable in a finite time under the influence of decoherence \cite{Song-PRA80-2009}. It has been shown explicitly that PPT-states are non-distillable (see Ref.\,\cite{Clarisse} and references therein). The quantum states with negative partial transpose (NPT) are regarded as distillable. There is a conjecture on the existence of NPT-bound entangled (NPT-non-distillable) states \cite{NPTBE-2000}, however the conclusive evidence is still missing. Therefore if an initial NPT-state becomes PPT-state in a finite time under the influence of dynamical process, then due to possibility of bound-entangled states, it suffers a transition from being distillable to non-distillable. This finite-time phase transition is called distillability sudden death (DSD).

Let us consider a particular family of quantum states given as
\begin{eqnarray}
\rho_\alpha = \frac{2}{7} \, |\Psi_+\rangle\langle\Psi_+| + \frac{\alpha}{7} \, \sigma_+ + \frac{5-\alpha}{7} \, \sigma_-\,, \label{Eq:H-BE}
\end{eqnarray}
where $2 \leq \alpha \leq 5$. In Eq.~(\ref{Eq:H-BE}) maximally entangled state $|\Psi_+\rangle = 1/\sqrt{3} (|0,1\rangle + |1,0\rangle + |2,2\rangle)$ is mixed with the separable states $\sigma_+ = 1/3 (| 0,0 \rangle\langle 0,0 | + |1,2 \rangle\langle 1,2 | + | 2,1 \rangle\langle 2,1 |)$ and $\sigma_- = 1/3 (| 1,1 \rangle\langle 1,1 | + |2,0 \rangle\langle 2,0 | + | 0,2 \rangle\langle 0,2 |)$. It was shown \cite{Horodecki-PRL82-1999} that $\rho_\alpha$ is separable for $2 \leq \alpha \leq 3$, bound-entangled for $3 < \alpha \leq 4$, and free-entangled for $4 < \alpha \leq 5$. Let us take these states with $4 < \alpha \leq 5$. The time evolution of these states can be easily determined from the general solution. In particular, we need to determine the time evolution of appropriate measure of entanglement for qutrit-qutrit systems. One appropriate measure for free-entangled states (NPT-states) is the negativity \cite{Vidal-PRA65-2002}, which is a measure of a state for having negative partial transpose. For positive value of this measure, the state will be NPT (hence entangled), whereas when this measure is zero, the state is PPT and we can not conclude the entanglement or separability of the state until some other measures/steps reveal the status of the state. The negativity is based on the partial transpose of a state $\rho$. The negativity $N(\rho)$ is equal to the absolute value of the sum of negative eigenvalues of partial transpose of a state $\rho$. Once the negativity becomes zero, we can easily study the time evolution of realignment criterion to determine the existence of bound-entangled states. However, we should keep in mind that realignment criterion can not detect all entangled states.

The partial transpose of the time evolved density matrix $\rho_\alpha(t)$ can have three possible negative eigenvalues. However, the expressions for possible negative eigenvalues are quite involved, therefore instead of providing them, we prefer to plot them for  specific choices of parameters $\alpha$, $\gamma_u$, and $\gamma_e$. The negativity for our example is given as
\begin{eqnarray}
N(\rho_\alpha(t)) = \max [0, -\eta_1] + \max[0,-\eta_2] + \max[0, - \eta_3]\,, \label{Eq:n-hbe}
\end{eqnarray}
where $\eta_i$ are the possible negative eigenvalues. 

\begin{figure}[h]
\scalebox{2.0}{\includegraphics[width=1.8in]{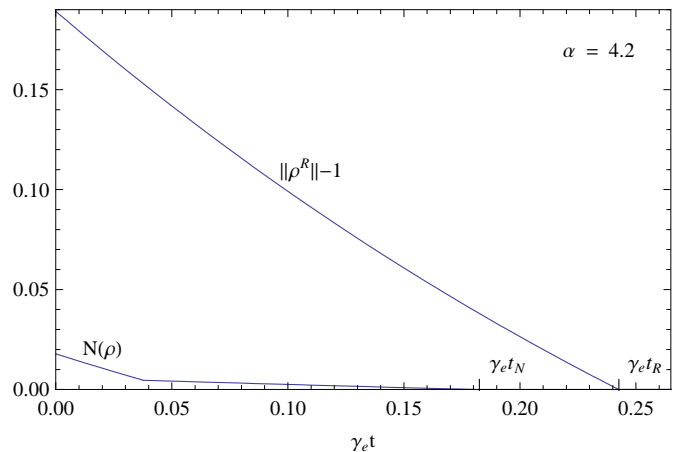}}
\centering
\caption{The negativity and realignment criterion is plotted against the decay parameter $\gamma_e t$ for $\alpha = 4.2$. We have defined $\gamma_u = 0.5 \, \gamma_e$.}
\label{Fig:dsd-hbe}
\end{figure} 

Let us now demonstrate that the time evolved density matrix $\rho_\alpha(t)$ do undergo distillability sudden death. This could be most easily demonstrated by checking entanglement of $\rho_\alpha(t)$ after it becomes PPT. Let us fix the interference parameter $\beta$ such that $\gamma_u = 0.5 \, \gamma_e$ throughout this paper which means that $\beta = 1/3$. The relation between negativity and realignment criterion is plotted in Figure \ref{Fig:dsd-hbe} against decay parameter $\gamma_e t$ and a specific choice of the single parameter $\alpha$. Figure \ref{Fig:dsd-hbe} shows that under amplitude damping an initial free-entangled state becomes bound-entangled at a time $\gamma_e t_N \approx 0.1826$. The entanglement of PPT-state is verified by the positive value of $\|\rho^R(t)\|-1 $ in the range $0.1826 \lesssim \gamma_e t \lesssim 0.2426$. However, the realignment criterion fails to detect the possible entanglement after time $\gamma_e t_R \approx 0.2426$. Hence we have demonstrated that free-entangled states exhibit distillability sudden death under amplitude damping.

\begin{figure}[h]
\scalebox{2.0}{\includegraphics[width=1.8in]{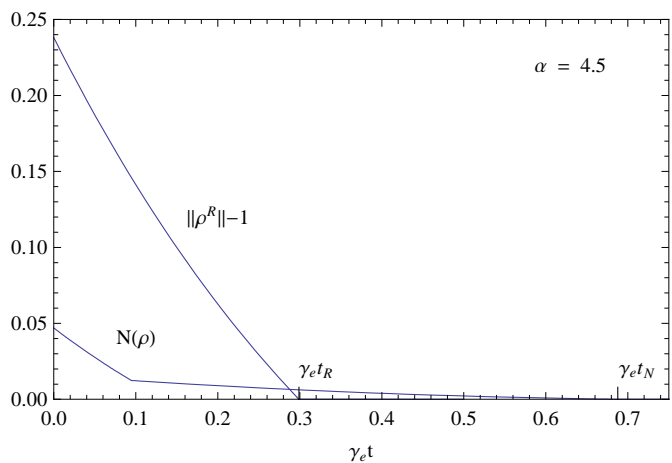}}
\centering
\caption{The negativity and realignment criterion is plotted against the decay parameter $\gamma_e t$ for $\alpha = 4.5$. We have taken $\gamma_u = 0.5 \, \gamma_e$.}
\label{Fig:dsd-hbep}
\end{figure}

Figure \ref{Fig:dsd-hbep} shows the time evolution of negativity and realignment criterion for an initial quantum state $\rho_\alpha$ (Eq.~(\ref{Eq:H-BE})) with $\alpha = 4.5$ and $\beta = 1/3$. For this particular case, the negativity becomes zero at $\gamma_e \, t_N \approx 0.6877$, however the realignment criterion fails to detect the entangled states after time $\gamma_e \, t_R \approx 0.299$. This example shows the sensitivity of realignment criterion on parameter $\alpha$. We have seen that although the initial NPT-states (free-entangled) become PPT after a finite time, however we can not conclude their separability/entanglement immediately. The PPT-states might be entangled suffering distillability sudden death followed by entanglement sudden death.

Let us now consider the locally equivalent state $\sigma_\alpha$, which may be obtained from Eq.~(\ref{Eq:H-BE}) by applying the local unitary operation $U = \mathbb{I}_3 \otimes \theta$, with $\theta = | 0 \rangle\langle 1 | + | 1 \rangle\langle 0 | + | 2 \rangle\langle 2 |$. Then we have 
\begin{eqnarray}
\sigma_\alpha = U \rho_\alpha U^\dagger = \frac{2}{7} \, |\tilde{\Psi}_+\rangle\langle \tilde{\Psi}_+| + \frac{\alpha}{7} \, \tilde{\sigma}_+ + \frac{5-\alpha}{7} \, \tilde{\sigma}_- \,. \label{Eq:lehbe}
\end{eqnarray}
As the local unitary transformations can not convert separable states into entangled states, therefore the transformed separable states are $\tilde{\sigma}_+ = 1/3 (| 0,1 \rangle\langle 0,1 | + |1,2 \rangle\langle 1,2 | + | 2,0 \rangle\langle 2,0 |)$, and $\tilde{\sigma}_- = 1/3 (| 1,0 \rangle\langle 1,0 | + |2,1 \rangle\langle 2,1 | + | 0,2 \rangle\langle 0,2 |)$. Similarly maximally entangled state is also converted into another maximally entangled state $|\tilde{\Psi}_+\rangle = 1/\sqrt{3} (|0,0\rangle + |1,1\rangle + |2,2\rangle)$. The local unitary transformations do not effect the trace and static entanglement of an initial quantum state, however they may have a profound influence on the future trajectory of entanglement \cite{Rau}. In the current situation the underlying dynamics is caused by spontaneous emission. For the range of $4 < \alpha \leq 5$, the dominant noisy component in Eq.~(\ref{Eq:H-BE}) is $\sigma_+$, which contains two doubly excited components $|1,2\rangle$, and $|2,1\rangle$, hence one could expect that entanglement and distillability may decay faster due to increase of probability of spontaneous emission. What we could achieve with local unitary transformations is to modify this component. Now the dominant noisy component in Eq.~(\ref{Eq:lehbe}) is $\tilde{\sigma}_+$ which contains only one doubly excited component $|1,2 \rangle$, therefore we expect considerable change in the dynamics of entanglement and distillability. Indeed, we find the expected intuitive results discussed below.

\begin{figure}[h]
\scalebox{2.0}{\includegraphics[width=1.8in]{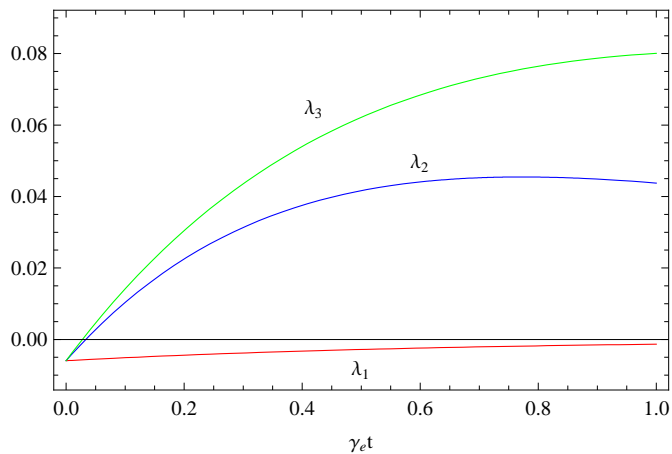}}
\centering
\caption{The possible negative eigenvalues are plotted against the decay parameter $\gamma_e t$ for $\alpha = 4.2$. The plot shows that the state remains NPT at all times except $\gamma_e t = \infty $.}
\label{Fig:dsd-lehbe1}
\end{figure}

The possible negative eigenvalues $\lambda_1$, $\lambda_2$, and $\lambda_3$ of the partial transpose of time evolved density matrix $\sigma_\alpha(t)$ are again quite involved, therefore instead of writing their expressions, we plot these eigenvalues against the decay parameter $\gamma_e t$. In Figure \ref{Fig:dsd-lehbe1}, we have plotted the negative eigenvalues against the decay parameter $\gamma_e t$. We observe that two of the negative eigenvalues become positive at some finite times, however one eigenvalue remains always negative in the range $4 < \alpha \leq 5$. This implies that the state $\sigma_\alpha$ remains always NPT-entangled until infinity, where it becomes separable. Hence distillability sudden death and entanglement sudden death never occur for $\sigma_\alpha$ under amplitude damping. In contrast the same state exhibits distillability sudden death and entanglement sudden death under multi-local dephasing \cite{Song-PRA80-2009, Ali-arXiv2009}. Hence we have seen that simple local unitary transformations can completely avoid distillability sudden death and entanglement sudden death under amplitude damping.

It is obvious from \ref{App} that for certain classes of initial states, our dynamical process evolves within that set. This simply means that all matrix elements which were initially zero remain zero. The examples studied in this paper are of such classes. Distillability sudden death has been observed so far (including this paper) for a very special class of states defined in Eq.~(\ref{Eq:H-BE}) and Eq.~(\ref{Eq:lehbe}) \cite{Song-PRA80-2009, Ali-arXiv2009}. These states are special in the sense that they already contain separable ($2 \leq \alpha \leq 3$), bound-entangled ($3 < \alpha \leq 4$) and free-entangled ($4 < \alpha \leq 5$) states. Our dynamical process decreases all matrix elements of initial states except $\rho_{99}(t)$. If we start with an initial free-entangled state, then it may happen that such decrement matches with bound-entangled region of this family causing distillability sudden death. Further dynamics bring all bound-entangled states to separable states.

Let us consider another important class of states called isotropic states \cite{Horodecki-PRA60-1999}. These states are given as
\begin{eqnarray}
\rho_p = p \, |\Psi_+\rangle \langle \Psi_+| + \frac{(1-p)}{9} \mathbb{I}_9 \,, 
\end{eqnarray}
where $0 \leq p \leq 1$. The isotropic states have the property that their PPT region is always separable (see Ref. \cite{Clarisse} and references therein). Therefore, these states do not suffer distillability sudden death under any type of decoherence, however they may suffer entanglement sudden death \cite{Derkacz-PRA74-2006}. This might indicate that distillability sudden death could be related to some particular classes of states which include bound-entangled as well as free-entangled states. 

\section{Discussions} \label{Sec:Conc}

We have studied the entanglement dynamics of bipartite systems under amplitude damping. In particular, we have focused the distillability sudden death of qutrit-qutrit systems. We have found that certain free-entangled qutrit-qutrit states become bound-entangled as a consequence of dissipative dynamical process. The explicit demonstration of distillability sudden death under the amplitude damping and phase damping suggests its generic nature. We have observed that local unitary transformations can avoid this phenomenon under amplitude damping. This situation is in contrast with phase damping where local unitary transformations can not avoid distillability sudden death. Therefore, we conclude that similar to multi-local dephasing case, qutrit-qutrit states interacting with statistically independent local reservoirs also evolve into three main types of dynamics: (i) They loose their entanglement only at infinity and hence never suffer entanglement sudden death and distillability sudden death. (ii) They loose their entanglement at a finite time but never undergo distillability sudden death. (iii) They first undergo distillability sudden death and then sudden death of entanglement.

\ack

I would like to thank Prof. Christof Wunderlich and his group for their kind hospitality at Universit\"at Siegen.

\appendix

\section{}\label{App}

The most general solution of the master equation (\ref{Eq:ME}) for an arbitrary initial state is given as
\begin{eqnarray*}
\rho_{11}(t) &=& \rho_{11} \, \mathrm{e}^{- 2 \, \gamma_e \, t} \, , \\ 
\rho_{12}(t) &=& \rho_{12} \, \mathrm{e}^{- \frac{(3 \, \gamma_e + \gamma_u) \, t}{2}} \, , \\ 
\rho_{13}(t) &=& \rho_{13} \, \mathrm{e}^{- \frac{3 \, \gamma_e \, t}{2}} \, , \\ 
\rho_{14}(t) &=& \rho_{14} \, \mathrm{e}^{- \frac{(3 \, \gamma_e + \gamma_u) \, t}{2}} \, , \\ 
\rho_{15}(t) &=& \rho_{15} \, \mathrm{e}^{- (\gamma_e + \gamma_u) \, t} \, , \\ 
\rho_{16}(t) &=& \rho_{16} \, \mathrm{e}^{- \frac{( 2 \, \gamma_e + \gamma_u) \, t}{2}} \, , \\ 
\rho_{17}(t) &=& \rho_{17} \, \mathrm{e}^{- \frac{3 \, \gamma_e \, t}{2}} \, , \\ 
\rho_{18}(t) &=& \rho_{18} \, \mathrm{e}^{- \frac{(2 \, \gamma_e +\gamma_u)\, t}{2}} \, , \\ 
\rho_{19}(t) &=& \rho_{19} \, \mathrm{e}^{- \gamma_e \, t} \, , \\ 
\rho_{22}(t) &=& \rho_{22} \, \mathrm{e}^{-(\gamma_e + \gamma_u) \, t} \, , \\ 
\rho_{23}(t) &=& \rho_{23} \, \mathrm{e}^{- \frac{(2 \, \gamma_e  + \gamma_u)\, t}{2}} \, , \\ 
\rho_{24}(t) &=& \rho_{24} \, \mathrm{e}^{- (\gamma_e + \gamma_u) \, t} \, , \\ 
\rho_{25}(t) &=& \rho_{25} \, \mathrm{e}^{- \frac{(\gamma_e + 3 \, \gamma_u) \, t}{2}} \, , \\ 
\rho_{26}(t) &=& \rho_{26} \, \mathrm{e}^{- \frac{(\gamma_e + 2 \, \gamma_u) \, t}{2}} \, , \\ 
\rho_{27}(t) &=& \rho_{27} \, \mathrm{e}^{- \frac{(2 \, \gamma_e + \gamma_u) \, t}{2}} \, , \\ 
\rho_{28}(t) &=& \rho_{28} \, \mathrm{e}^{- \frac{(\gamma_e + 2 \, \gamma_u) \, t}{2}} \, , \\ 
\rho_{29}(t) &=& \rho_{29} \, \mathrm{e}^{- \frac{(\gamma_e +\gamma_u) \, t}{2}} \, , \\ 
\rho_{33}(t) &=& [\rho_{11} + \rho_{22} + \rho_{33}] \, \mathrm{e}^{- \gamma_e \, t} - \rho_{11}(t) -\rho_{22}(t) \, , \\ 
\rho_{34}(t) &=& \rho_{34} \, \mathrm{e}^{- \frac{(2 \, \gamma_e + \gamma_u) \, t}{2}} \, , \\ 
\rho_{35}(t) &=& \rho_{35} \, \mathrm{e}^{- \frac{(\gamma_e + 2 \, \gamma_u) \, t}{2}} \, , \\ 
\rho_{36}(t) &=& [\rho_{14} + \rho_{25} + \rho_{36}] \, \mathrm{e}^{- \frac{(\gamma_e +\gamma_u) \, t}{2}} - \rho_{14}(t) - \rho_{25}(t) \, , \\ \rho_{37}(t) &=& \rho_{37} \, \mathrm{e}^{- \gamma_e \, t} \, , \\ 
\rho_{38}(t) &=& \rho_{38} \, \mathrm{e}^{- \frac{(\gamma_e +\gamma_u) \, t}{2}} \, , 
\end{eqnarray*}
\begin{eqnarray*} 
\rho_{39}(t) &=& [\rho_{17} + \rho_{28} + \rho_{39}] \, \mathrm{e}^{- \frac{\gamma_e  \, t}{2}} - \rho_{17}(t) - \rho_{28}(t) \, , \\ 
\rho_{44}(t) &=& \rho_{44} \, \mathrm{e}^{- (\gamma_e +\gamma_u) \, t} \, , \\ 
\rho_{45}(t) &=& \rho_{45} \, \mathrm{e}^{- \frac{(\gamma_e + 3 \, \gamma_u) \, t}{2}} \, , \\ 
\rho_{46}(t) &=& \rho_{46} \, \mathrm{e}^{- \frac{(\gamma_e + 2 \, \gamma_u) \, t}{2}} \, , \\  
\rho_{47}(t) &=& \rho_{47} \, \mathrm{e}^{- \frac{(2 \, \gamma_e + \gamma_u) \, t}{2}} \, , \\ 
\rho_{48}(t) &=& \rho_{48} \, \mathrm{e}^{- \frac{(\gamma_e + 2 \, \gamma_u) \, t}{2}} \, , \\ 
\rho_{49}(t) &=& \rho_{49} \, \mathrm{e}^{- \frac{(\gamma_e + \gamma_u) \, t}{2}} \, , \\
\rho_{55}(t) &=& \rho_{55} \, \mathrm{e}^{- 2 \, \gamma_u \, t} \, , \\ 
\rho_{56}(t) &=& \rho_{56} \, \mathrm{e}^{- \frac{3 \, \gamma_u \, t}{2}} \, , \\ 
\rho_{57}(t) &=& \rho_{57} \, \mathrm{e}^{- \frac{(\gamma_e + 2 \, \gamma_u) \, t}{2}} \, , \\ 
\rho_{58}(t) &=& \rho_{58} \, \mathrm{e}^{- \frac{3 \, \gamma_u \, t}{2}} \, , \\ 
\rho_{59}(t) &=& \rho_{59} \, \mathrm{e}^{- \gamma_u \, t} \, , \\ 
\rho_{66}(t) &=& [\rho_{44} + \rho_{55} + \rho_{66}] \, \mathrm{e}^{- \gamma_u \, t} - \rho_{44}(t) - \rho_{55}(t) \, , \\ 
\rho_{67}(t) &=& \rho_{67} \, \mathrm{e}^{- \frac{(\gamma_e + \gamma_u) \, t}{2}} \, , \\ 
\rho_{68}(t) &=& \rho_{68} \, \mathrm{e}^{- \gamma_u \, t} \, , \\
\rho_{69}(t) &=& [\rho_{47} + \rho_{58} + \rho_{69}] \, \mathrm{e}^{- \frac{\gamma_u \, t}{2}} - \rho_{47}(t) - \rho_{58}(t) \, , \\ 
\rho_{77}(t) &=& [\rho_{11} + \rho_{44} + \rho_{77}] \, \mathrm{e}^{- \gamma_e \, t} - \rho_{11}(t) - \rho_{44}(t) \, , \\ 
\rho_{78}(t) &=& [\rho_{12} + \rho_{45} + \rho_{78}] \, \mathrm{e}^{- \frac{(\gamma_e + \gamma_u)\, t}{2}} - \rho_{12}(t) - \rho_{45}(t) \, , \\ \rho_{79}(t) &=& [\rho_{13} + \rho_{46} + \rho_{79}] \, \mathrm{e}^{- \frac{ \gamma_e \, t}{2}} - \rho_{13}(t) - \rho_{46}(t) \, , \\ 
\rho_{88}(t) &=& [\rho_{22} + \rho_{55} + \rho_{88}] \, \mathrm{e}^{- \gamma_u \, t} - \rho_{22}(t) - \rho_{55}(t) \, , \\ 
\rho_{89}(t) &=& [\rho_{23} + \rho_{56} + \rho_{89}] \, \mathrm{e}^{- \frac{ \gamma_u \, t}{2}} - \rho_{23}(t) - \rho_{56}(t) \, , \\
\rho_{99}(t) &=& 1 + \Theta(t) 
\end{eqnarray*} 
where $\Theta(t) = \mathrm{e}^{-2 (\gamma_e + \gamma_u) t} [ \, \rho_{11} \, \mathrm{e}^{2 \, \gamma_u \, t} + (\rho_{22} + \rho_{44}) \mathrm{e}^{(\gamma_e + \gamma_u) t} + \rho_{55} \, \mathrm{e}^{2 \, \gamma_e \, t} - (2 \rho_{11} + \rho_{22} + \rho_{33} + \rho_{44} + \rho_{77} ) \mathrm{e}^{(\gamma_e + 2 \gamma_u) t} - (\rho_{22} + \rho_{44} + 2 \rho_{55} + \rho_{66} + \rho_{88} ) \mathrm{e}^{(2 \gamma_e + \gamma_u) t} \, ]$.

\end{document}